\documentclass{sig-alternate}
\usepackage{amssymb}
\usepackage{graphicx}
\usepackage{hyperref}
\hypersetup{colorlinks=false}

\usepackage{algorithm}
\usepackage{algorithmic}

\usepackage{amsmath}
\usepackage{amsfonts}
\usepackage{amssymb}

\newtheorem{theorem}{Theorem}[section]
\newtheorem{lem}[theorem]{Lemma}

\newtheorem{thm}[theorem]{Theorem}
\newtheorem{cor}[theorem]{Corollary}
\newdef{rem}{Remark}
\newdef{exmp}{Example}
\newtheorem{defn}[theorem]{Definition}
\numberwithin{theorem}{section} \numberwithin{equation}{section}

\begin{document}

\title{ Simultaneous Integer Relation Detection and\\
Its an Application}

\numberofauthors{1}
\author{
\alignauthor Chen Jing-wei $^{^{^{^{\S \,\P}}}}$,\ Feng
Yong\titlenote{Corresponding author.}$^{^{^{^{\S}}}}$,\ Qin
Xiao-lin$^{^{^{^{\S \,\P}}}}$\
and \ \ Zhang Jing-zhong$^{^{^{^{\S\,\P }}}}$\\
       \affaddr{\ }\\
       \affaddr{{$^{^{^{{\S}}}}$ Laboratory of Computer Reasoning and
        Trustworthy Computation,}}\\
       \affaddr{{University of Electronic Science and Technology of China, Chengdu 610054, China}}\\
       \affaddr{{$^{^{^{{\P}}}}$Laboratory for Automated Reasoning and
       Programming, Chengdu Institute of} }\\
       \affaddr{{Computer Applications, Chinese Academy of Sciences, Chengdu 610041, China}}\\
      \email{
         {{\small \texttt{velen.chan@163.com\ \ \ \ $\{$yongfeng,$\,$qinxl$\}\,$@casit.ac.cn\ \ \ \
         zjz101@yahoo.com.cn}}}}
} \maketitle

\begin{abstract}
Let $\mathbf{x_1}, \cdots, \mathbf{x_t} \in  \mathbb{R}^{n}$. A
simultaneous integer relation (SIR) for $\mathbf{x_1}, \cdots,
\mathbf{x_t}$ is a vector $\mathbf{m} \in
\mathbb{Z}^{n}\setminus\{\textbf{0}\}$ such that
$\mathbf{x_i}^T\mathbf{m} = 0$ for $i = 1, \cdots, t$. In this
paper, we propose an algorithm SIRD to detect an SIR for real
vectors, which constructs an SIR within $\mathcal {O}(n^4 + n^3 \log
\lambda(X))$ arithmetic operations, where $\lambda(X)$ is the least
Euclidean norm of SIRs for $\mathbf{x_1}, \cdots, \mathbf{x_t}$. One
can easily generalize SIRD to complex number field. Experimental
results show that SIRD is practical and better than another
detecting algorithm in the literature. In its application, we
present a new algorithm for finding the minimal polynomial of an
arbitrary complex algebraic number from its an approximation, which
is not based on LLL. We also provide a sufficient condition on the
precision of the approximate value, which depends only on the height
and the degree of  the
algebraic number.
\end{abstract}


\section{Introduction}
\label{sec:introduction}

Let
 $\mathbf{x_1}, \cdots, \mathbf{x_t}$ be vectors in $\mathbb{R}^n$,
and denote $(\mathbf{x_1}, \cdots, \mathbf{x_t})$ by $X$. A
\emph{simultaneous integer relation} (SIR) for $\mathbf{x_1},
\cdots, \mathbf{x_t}$ is a vector $\mathbf{m} \in
\mathbb{Z}^{n}\setminus\{\textbf{0}\}$ such that $X^T\mathbf{m} =
\textbf{0}$, i.e. $\mathbf{x_i}^T\mathbf{m} = 0$ for $i = 1,\cdots,
t$. For short, we also call $\mathbf{m}$  an SIR for $X$. When $t =
1$, we say that $\mathbf{m}$ is an integer relation for
$\mathbf{x_1}$.
The problem of detecting integer relations for a 
rational or real vector is quite old.
Historical surveys 
can be found in \cite{Ber1971, FF1979,
Bre1982, HJL1989, FBA1999}. Among these integer relation detecting
algorithms, the HJLS algorithm \cite{HHL1986, HJL1989} and the PSLQ
algorithm \cite{FB1992, FBA1999} have been used frequently.

In the present paper, using  the technique to construct the
hyperplane matrix in HJLS and a generalized method of the matrix
reduction from PSLQ we propose an 
algorithm SIRD, which can be
used to detect an SIR for $t$ real vectors. 
The cost of our algorithm is at most $\mathcal {O}(n^4 + n^3 \log
\lambda(X))$ exact arithmetic operations for detecting an SIR for
$X$, where $\lambda(X)$ represents the least Euclidean norm of SIRs
for $X$. Furthermore, our detecting algorithm SIRD  either always
finds an SIR for $X$ if one exists or proves that there are no SIRs
for $X$ of norm less than a given size. Experimental results show
that SIRD is practical.

In application, we successfully apply  SIRD to find the minimal
polynomial of an algebraic number $\alpha \in \mathbb{C}$ with
degree and height at most $n$ and $H$ respectively from its an
approximation $\bar{\alpha}$ satisfying $\max_{1\leq i\leq
n}|\alpha^{i}-\bar{\alpha}^i| < \epsilon$, and propose the
corresponding 
algorithm MPF, where  the
minimal polynomial of an algebraic number $\alpha$ is the unique
primitive polynomial $p(x) \in \mathbb{Z}[x]$ of least degree such
that $p(\alpha)  =
 0$.
In fact, for $i$ from $1$ to $n$ we run  SIRD with
$\mathbf{v_1}=(1,\mbox{Re}(\bar{\alpha}),\cdots,\mbox{Re}(\bar{\alpha}^i))^T$,
$\mathbf{v_2} = (0, \mbox{Im}(\bar{\alpha}), \cdots,
\mbox{Im}(\bar{\alpha}^i))^T$ as its input and then an exact SIR for
$\mathbf{v_1}, \mathbf{v_2}$ has been detected. 
We provide  a sufficient controlling on $\epsilon$ and prove that
such an $\epsilon$ is sufficient to enable an exact SIR for
$\mathbf{v_1}$ and $\mathbf{v_2}$ to be also an SIR for
$(1,\mbox{Re}(\alpha),\cdots,\mbox{Re}(\alpha^i))^T$ and $(0,
\mbox{Im}(\alpha), \cdots, \mbox{Im}(\alpha^i))^T$,
where 
$\epsilon$ depends only on $n$ and $H$, as in
(\ref{eq:error-controlling}). It implies the correctness of MPF and
is better than already existing results in \cite{Jus1989,QFC2009}.
\subsection{Related Works}
In \cite{HHL1986, HJL1989}, J. Hastad, B. Just, J. C. Lagarias, and
C. P. Schnorr not only presented the HJLS algorithm and the first
rigorous proof of a `polynomial time' bound for a relation finding
algorithm but also proposed a simultaneous relations algorithm (see
 \cite[section 5]{HJL1989}), whereas HJLS
 is numerically unstable. The unstable examples can be
found in \cite{FB1992, FBA1999}. 
 In their draft \cite{RS1997}, C. R{\"o}ssner and C. P.
Schnorr studied the case of $t = 2$ by using a modified HJLS
algorithm. But for the moment, \cite{RS1997} is still in a
preliminary state with some open problems. The PSLQ algorithm,
together with related lattice reduction schemes such as LLL
\cite{LLL1982}, was named one of ten ``algorithms of the twentieth
century'' by the publication \emph{Computing in Science and
Engineering} (see \cite{DS2000, BBC2007}), and is now extensively
used in Experimental Mathematics, with applications such as
identification of multiple zeta constants,
a new formula for $\pi$, finding algebraic relations 
 and so on (see \cite{BB2001, BBC2007, BB2009}). Moreover,
PSLQ is numerically stable and can be easily generalized to complex
number field and Hamiltonian quaternion number field (see
\cite{FBA1999}), but
 it is not suitable to
detect an SIR for several real vectors.
\\

The SIRD algorithm in this paper is to detect an SIR for $t$ real
vectors and can be applied to detect an integer relation in
$\mathbb{Z}^n$ for a
 complex vector or a Hamilton quaternion number vector. A significant
body of experimental data shows that SIRD is practical and better
than the HJLS simultaneous relations algorithm.


In fact, the MPF algorithm in this paper is a  positive answer to
the following interesting question: Suppose we are given an
approximation to an algebraic number $\alpha$, and two bounds on the
degree and  the size of the coefficients of its minimal polynomial
respectively. Is it possible to infer the minimal polynomial? The
question was raised, independently, by Manuel Blum  in theoretical
cryptography (see \cite{KLL1984,KLL1988}) and the last author of
this paper  in automated reasoning (see \cite{YZH1996}). The first
complete answer to this question, KLL algorithm, was presented by R.
Kannan, A.K. Lenstra and L. Lov{\'a}sz in \cite{KLL1984, KLL1988} by
using the celebrated lattice reduction algorithm LLL \cite{LLL1982}.
In the computer algebra system \emph{Maple}, the built-in function
\texttt{PolynomialTools:-MinimalPolynomial()} is a function to find
a polynomial of degree $n$ (or less) with small integer coefficients
which has the given approximation $r$ of an algebraic number as one
of its roots and is based on  KLL algorithm. The correctness of the
polynomial returned by the built-in function
depends on the accuracy of the approximation (see \emph{Maple}'s Help). 
 From another aspect, the minimal
polynomial of an algebraic number $\alpha$ with exact degree $n$ can
be found by detecting an integer relation for the vector $\mathbf{v}
= (1, \alpha, \cdots, \alpha^n)^T$. Besides
HJLS, 
B. Just also presented an algorithm to detect integer relations for
a given vector consists of algebraic numbers in \cite{Jus1989}. We
can apply Just's algorithm or  HJLS  to the vector $\mathbf{v}$ for
finding the minimal polynomial of $\alpha$.
However, 
both Just's algorithm and HJLS 
are not numerically stable, as mentioned previously. All these
algorithms are based on LLL. Two authors of this paper presented a
method to reconstruct a rational number from its an approximation by
using continued fraction in \cite{ZF2007}. It may be viewed as an
answer to a special case of  the question.
Based on PSLQ, one can find algebraic relations, 
such as \cite{BC1999, BL2000, BHM2002, BBK2006}, whereas these
articles did not involve the minimal polynomial finding. The authors
of this paper also presented an algorithm in \cite{QFC2009} for
finding the minimal polynomial of a real algebraic number from its
an approximation. However, these PSLQ based algorithms can not deal
with complex algebraic numbers since PSLQ only outputs  a relation
in Gaussian integer ring
 for a complex vector. 

Fortunately, our simultaneous integer relation detection algorithm
SIRD in present paper can be used to overcome these pitfalls.
Applying SIRD to one or two real vectors, we present another
affirmative answer, the MPF algorithm, to the question above. We
show that MPF is a more efficient minimal polynomial finding
algorithm comparing with the algorithms in \cite{Jus1989,QFC2009}
and provide a sufficient condition on the error controlling, from
which we can claim that the
polynomial returned by 
MPF is the exact minimal polynomial of the algebraic number that we
only know an approximate value and two bounds on its degree and
height. Although a  similar even better complexity can be obtained
by KLL, MPF has its own meaning 
since it is a new method without using LLL reduction.

\textbf{{Road-map.}} In section \ref{sec:The Simultaneous Integer
Relation Detecting Algorithm} and \ref{sec:The SIRD algorithm} we
first give some preliminaries, and then present the SIRD algorithm
and analyze it. We report on some experimental results about the
performance of SIRD in section \ref{sec:Performance results}, apply
SIRD to find the minimal polynomial of an algebraic number from its
an approximation and propose the MPF algorithm in section
\ref{sec:finding-minimal-polynomial}, in which we also analyze  MPF
and present the result of error controlling. We conclude this paper
with section \ref{sec:conclude}.

\textbf{{Notations.}} Throughout this paper, $\mathbb{Z}$,
$\mathbb{R}$, and $\mathbb{C}$ are the sets of integers,  real
numbers, and complex numbers respectively. The real and imaginary
parts of $z\in\mathbb{C}$ will be denoted Re$(z)$ and Im$(z)$
respectively. For $c\in\mathbb{R}$, 
$\lfloor c \rceil = \lfloor c+\frac{1}{2}\rfloor$. All vectors in
this paper are column vectors, and will be denoted in bold.  If
$\mathbf{x} \in \mathbb{R}^n$, then ~$\|\mathbf{x}\|_2$~represents
its Euclidean norm, i.e. $\|\mathbf{x}\|_2 = \sqrt{<\mathbf{x},
\mathbf{x}>}$, where~$<*, *>$ is the inner product of two vectors.
We denote $n\times n$ identity matrix by $I_{n}$. Given a matrix $A
= (a_{i,j})$, we denote its transpose by $A^T$, its trace by
tr$(A)$, its determinant by $|A|$, and its Frobenius norm by
 $\|A\|_F = (\mbox{tr}(A^TA))^{1/2}$, i.e. $\|A\|_F = (\sum
a_{i,j}^2)^{1/2}$. We say that a matrix $A$ is lower trapezoidal if
$a_{i,j} = 0$ for $i < j$. $GL(n, \mathbb{Z})$ is the group of
$n\times n$ unimodular matrix with entries in $\mathbb{Z}$. The
height of a vector is defined by the maximum of all the absolute
values of its entries. For a polynomial $f(x) = \sum_{i=0}^n
f_ix^i$, we denote by $\deg(f)$ its degree with respect to $x$,
$\|f\|_{1} = \sum_{i=0}^n|f_{i}|$ its one norm,
 $\|f\|_2 = (\sum_{i=0}^n|f_{i}|^{2})^{1/2}$ its Euclidean length, and
height$(f)=\max_{0 \leq i \leq n}|f_{i}|$ its height.

\section{Preliminaries}
\label{sec:The Simultaneous Integer Relation Detecting Algorithm}

In what follows we always suppose that $\mathbf{x_1}, \cdots,
\mathbf{x_t}$ are linearly independent vectors in $\mathbb{R}^n$,
where $\mathbf{x_i} = (x_{i,1}, \cdots, x_{i,n})^T$. Obviously, we
have $t < n$. We denote by $X$ the matrix $(\mathbf{x_1}, \cdots,
\mathbf{x_t})$, and suppose that $X \in \mathbb{R}^{n\times t}$
satisfies
\begin{equation}\label{eq:suppose-on-x} \left|
  \begin{array}{cccc}
    x_{1,n-t+1} & x_{2,n-t+1} & \cdots & x_{t,n-t+1} \\
    x_{1,n-t+2} & x_{2,n-t+2}  & \cdots & x_{t,n-t+2}\\
    \vdots &\vdots& &\vdots\\
    x_{1,n} & x_{2,n}  & \cdots & x_{t,n}
  \end{array}
\right| \neq 0,\end{equation} unless otherwise specified. For $X \in
\mathbb{R}^{n\times t}$ not satisfying (\ref{eq:suppose-on-x}),
exchanging some rows of $X$ produces $X' = CX$, where $C$ is an
appropriate matrix in $GL(n, \mathbb{Z})$. And then we detect an SIR
for $X'$. If $\mathbf{m}$ is an SIR for $X'$, then $C^T\mathbf{m}$
is an SIR for $X$.
\subsection{Hyperplane Matrix}\label{subsec:Hyperplane-matrix}

\begin{defn}[Hyperplane Matrix]\label{def:Hyperplane-matrix}
Let~$X=(\mathbf{x_1},\\\cdots, \mathbf{x_t}) \in \mathbb{R}^{n\times
t}$. A hyperplane matrix  with respect to $X$ is any matrix $W \in
\mathbb{R}^{n\times(n-t)}$
 such that $X^TW = \mathbf{0}$ and the columns of $W$ $\mbox{span}$
$X^{\perp} = \{ \mathbf{y}\in\mathbb{R}^n: \mathbf{x_i}^T\mathbf{y}
= 0, i = 1, \cdots, t\}$.
\end{defn}

Now we introduce a method to construct a hyperplane matrix for  $X$.

Let $\mathbf{b_1}, \cdots, \mathbf{b_n}$~form a standard basis of
~$\mathbb{R}^{n}$, i.e. the $i$-th entry of $\mathbf{b_i}$ is $1$
and others are $0$. By performing the process of standard
Gram-Schmidt orthogonalization to $\mathbf{x_1},\,\cdots,\,\mathbf{x_t},\\\mathbf{b_1},\cdots,\mathbf{b_n}$ 
in turn we have
\[
\begin{split}
&\mathbf{x_1}^* =
\frac{\mathbf{x_1}}{\|\mathbf{x_1}\|_2},\,\mathbf{x_k}^* =
\mathbf{x_k} - \sum_{j=1}^{k-1}\frac{<\mathbf{x_k},
\mathbf{x_j}^*>}{<\mathbf{x_j}^*,
\mathbf{x_j}^*>}\mathbf{x_j}^*,\\&\mathbf{x_k}^* =
\frac{\mathbf{x_k}^*}{\|\mathbf{x_k}^*\|_2}, k = 2, \cdots,
t,\\&\mathbf{b_1}^* = \mathbf{b_1} -
\sum_{j=1}^{t}\frac{<\mathbf{b_1}, \mathbf{x_j}^*>}{<\mathbf{x_j}^*,
\mathbf{x_j}^*>}\mathbf{x_j}^*, \mathbf{b_1}^*
 = \frac{\mathbf{b_1}^*}{\|\mathbf{b_1}^*\|_2},\\\end{split}\]\[\begin{split}
 &\mathbf{b_i}^* = \mathbf{b_i} -
\sum_{j=1}^{t}\frac{<\mathbf{b_i}, \mathbf{x_j}^*>}{<\mathbf{x_j}^*,
\mathbf{x_j}^*>}\mathbf{x_j}^*
 - \sum_{j=1}^{i-1}\mu_{i,j}\mathbf{b_j}^*
,        \\&\mathbf{b_i}^* =
\frac{\mathbf{b_i}^*}{\|\mathbf{b_i}^*\|_2},        i = 2, \cdots,
n,
\end{split}
\]
where
$
\mu_{i,j} = \left\{
\begin{aligned}
         &\frac{<\mathbf{b_i}, \mathbf{b_j}^*>}{<\mathbf{b_j}^*, \mathbf{b_j}^*>}, &\mbox{if} \ \ \|\mathbf{b_j}^*\|_2 \neq 0,\\
                  &0,
                   &\mbox{if}  \ \  \|\mathbf{b_j}^*\|_2 = 0.
                                            \end{aligned} \right.
$

\begin{lem}\label{lem:hyperplane-matrix-construction}
Let $\mathbf{x_k}$, $\mathbf{x_k}^*$, $\mathbf{b_j}$ and
$\mathbf{b_j}^*$ be as above. Then

1. there exist $t$ elements in $\{1, \cdots, n\}$ denoted by $j_1
\cdots,j_t$ such that
 $\mathbf{b_{j_1}}^* = \cdots = \mathbf{b_{j_t}}^* = \textbf{0}$.

2. $\mathbf{b_{n-t+1}}^* = \cdots = \mathbf{b_{n}}^* = \textbf{0}$.
\end{lem}

\begin{proof}
Part $1$ easily follows from the process of standard Gram-Schmidt
orthogonalization. We next prove $\mathbf{b_{n-t+1}}^* = \cdots =
\mathbf{b_{n}}^* = \textbf{0}$ when  (\ref{eq:suppose-on-x}) holds.
Set
\[
a_1\mathbf{x_1} + \cdots + a_t\mathbf{x_t} + l_1\mathbf{b_1} +
\cdots + l_{n-t}\mathbf{b_{n-t}} = \textbf{0}.
\]
Taking each side as a column vector and observing the last $t$
components of two sides, we have $a_1 = \cdots = a_t = 0$. And since
$\mathbf{b_1}, \cdots, \mathbf{b_{n-t}}$ are linearly independent,
we have $l_1 = \cdots = l_{n-t} = 0$. Thus the $n$ vectors
$\mathbf{x_1}, \cdots, \mathbf{x_t}, \mathbf{b_1}, \cdots,
\mathbf{b_{n-t}}$ are linearly independent. This implies that
$\mathbf{b_{n-t+1}}^* = \cdots = \mathbf{b_{n}}^* = \textbf{0}$.
\end{proof}


\begin{defn}[$H_X$]
\label{def:H_X} For $X\in\mathbb{R}^{n\times t}$ satisfying
(\ref{eq:suppose-on-x}), 
define $H_{X}$ to be the $n\times (n-t)$ matrix $(\mathbf{b_1}^*,
\cdots, \mathbf{b_{n-t}}^*)$.
\end{defn}

\begin{lem}\label{lem:property-of-Hx}
Let $X\in\mathbb{R}^{n\times t}$ and $H_{X}$ be as above. Then

1. $H_{X}^T H_{X} = I_{n-t}$.

2. $\|H_{X}\|_F = \sqrt{n-t}$.

3. $(\mathbf{x_1}^*, \cdots, \mathbf{x_t}^*, H_{X})$ is an
orthogonal matrix.

4. $X^T H_{X} = $\textbf{0}, i.e. $H_{X}$ is a hyperplane matrix of
$X$.


5. $H_{X}$ is a lower trapezoidal matrix and every diagonal element
of $H_{X}$ is nonzero.
\end{lem}

\begin{proof}
Since every two columns of $H_{X}$ are orthogonal, part $1$ follows.
And  part $2$ follows from part $1$. Let $X^* = (\mathbf{x_1}^*,
\cdots, \mathbf{x_t}^*)^T$. Obviously,
$(\mathbf{x_1}^*,\cdots,\mathbf{x_t}^*,H_{X})$ is an orthogonal
matrix. From part 3 and standard Gram-Schmidt orthogonalization we
have $X^{*T}H_{X} = \textbf{0}$ and $X = X^*Q$ respectively, where
$Q$ is an appropriate $t\times t$ invertible matrix. Thus $X^TH_{X}
= Q^TX^{*T}H_{X}= \textbf{0}$ and hence that part 4 follows. We now
prove part 5. Denote the $k$-th element of $\mathbf{b_i}^*$ by
$b_{i,k}^*$. The diagonal elements of $H_{X}$ are $b_{i,i}^{*}$ for
$i = 1, \cdots, n-t$. Before normalizing $\mathbf{b_i}^*$ we have $
b_{i,i}^* = 1 - \sum_{k=1}^{t}x_{k,i}^{*2} -
\sum_{j=1}^{i-1}b_{j,i}^{*2}$, and at the same time, $ 0 \neq
\|\mathbf{b_{i}}^*\|_2^2 = <\mathbf{b_{i}}^*, \mathbf{b_{i}}^*>
 = 1 - \sum_{k=1}^{t}x_{k,i}^{*2} - \sum_{j=1}^{i-1}b_{j,i}^{*2}.
$ Thus all the diagonal elements of $H_{X}$ are nonzero. Now we only
need to show that $H_{X}$ is lower trapezoidal. From standard
Gram-Schmidt orthogonalization, we can check that $b_{i,k}^* =
<\mathbf{b_i}^*, \mathbf{b_k}^*> = 0$ holds for $i> k$. This
completes the proof.
\end{proof}

So far, we have had a method to produce a hyperplane matrix $H_X$
for $X \in \mathbb{R}^{n\times t}$. The basic idea is from HJLS
(see \cite{HHL1986, HJL1989}). The same strategy was also used in
PSLQ, however, in which partial sum was adopted instead of
Gram-Schmidt orthogonalization.

\begin{lem}\label{lem:property-of-Px}
For $X = (\mathbf{x_1}, \cdots, \mathbf{x_t})\in\mathbb{R}^{n\times
t}$ define $P_{X} = H_{X} H_{X}^T$. Then

1. $P_{X}^T = P_{X}$.

2. $P_{X} = I_n - \sum_{i=1}^t\mathbf{x_i}^* \mathbf{x_i}^{*T}$.

3. $P_{X}^2 = P_{X}$.

4. $\|P_{X}\|_F = \sqrt{n-t}$.

5. $P_{X} \mathbf{z} = \mathbf{z}$ for any $\mathbf{z} \in X^\perp$.
Particularly, $P_{X} \mathbf{m} = \mathbf{m}$ for any SIR
$\mathbf{m}$ for $X$.
\end{lem}

\begin{proof}
The proof of the first part is easy. Let~$U = (\mathbf{x_1}^*,
\cdots, \mathbf{x_t}^*, H_{X})$. From Lemma \ref{lem:property-of-Hx}
we have
 $I_n=UU^T= H_{X} H_{X}^T
 + \sum_{i=1}^t\mathbf{x_i}^* \mathbf{x_i}^{*T}$. Thus part $2$
follows. Part $3$ and part $4$ follow from
 $P_{\textbf{x}}^2 = H_{X}(H_{X}^TH_{X})H_{X}^T = H_{X}H_{X}^T = P_{X}$
and $\|P_{X}\|_F^2 =$ tr$(P_{X}^T P_{X}) =$ tr$(P_{X}) = $
tr$(H_{X}^T H_{X}) = n - t$ respectively. Since $\mathbf{z} \in
X^\perp$, we have $<\mathbf{x_i}, \mathbf{z}> = 0$ for $i = 1
\cdots, t$. And the process of standard Gram-Schmidt
orthogonalization implies $\mathbf{x_i}^{*T} \mathbf{z} = 0$. Thus
we have $P_{X} \mathbf{z} = \mathbf{z} - (\sum_{i=1}^t\mathbf{x_i}^*
\mathbf{x_i}^{*T})\mathbf{z} = \mathbf{z}$ from part 2.
\end{proof}

From Lemma \ref{lem:property-of-Hx} and Lemma
\ref{lem:property-of-Px} we can easily generalize the Theorem $1$ in
\cite{FBA1999} to the case of $X \in \mathbb{R}^{n\times t}$.

\begin{thm}\label{thm:lower-bound-of-simultaneous-relations}
Let $X \in \mathbb{R}^{n\times t}$ and $H_{X}$ be as above. Suppose
that for any matrix $A \in GL(n, \mathbb{Z})$ there exists an
orthogonal matrix $Q \in \mathbb{R}^{(n-t)\times(n-t)}$ such that
$(h_{i,j}) = AH_{X}Q$ is lower trapezoidal and all of the diagonal
elements of $(h_{i,j})$ satisfy $h_{j,j} \neq 0$. Then for any SIR
$\mathbf{m}$ of $X$ we have
\begin{equation}\label{eq:lower-bound-of-simultaneous-relation}
\frac{1}{\max_{1\leq j\leq n-t}|h_{j,j}|} = \min_{1\leq j\leq
n-t}\frac{1}{|h_{j,j}|} \leq \|\mathbf{m}\|_2.
\end{equation}
\end{thm}

As this theorem 
easily follows from the proof of Theorem 1 of \cite{FBA1999} with
little modifications, the detail has been  omitted here.

 The lower
bound given in (\ref{eq:lower-bound-of-simultaneous-relation}) when
$t=1$ is consistent with a similar lower bound in \cite{FF1979,
FF1982}. Moreover, if a method to reduce the norm of $H_X$ by
multiplication by some unimodular $A\in GL(n, \mathbb{Z})$ on the
left has been developed, then it will produce an increasing lower
bound on $\lambda(X)$, where $\lambda(X)$ is the least Euclidean
norm of SIRs for $X$. In fact this theorem suggests a strategy to
detect an SIR for $X$.


\subsection{Matrix Reduction}\label{subsec:Matrix-Reudce}
We now study how to reduce the hyperplane matrix $H_X$. 
First we recall (modified) Hermite reduction 
in \cite{FBA1999}.


\begin{algorithm}[H]
\caption{(Modified Hermite Reduction).}
\begin{algorithmic}[1]
\REQUIRE a lower trapezoidal matrix $H=(h_{i,j})\in
\mathbb{R}^{n\times (n-1)}$ with $h_{j,j} \neq 0$.

\ENSURE a reducing matrix $D$ of $H$.

\STATE {$D := I_n$}

\STATE {\textbf{for} $i$ from $2$ to $n$ \textbf{do}}

\STATE {\ \ \ \ \textbf{for} $j$ from $i-1$ by $-1$ to $1$
\textbf{do}}

\STATE {\ \ \ \ \ \ \ \ $q := \lfloor h_{i,j}/h_{j,j}\rceil$, where
$\lfloor c\rceil = \lfloor c + 1/2\rfloor$ for a real\\\ \ \ \ \ \ \
\ number $c$.}

\STATE {\ \ \ \ \ \ \ \ \textbf{for} $k$ from $1$ to $n$
\textbf{do}}

\STATE {\ \ \ \ \ \ \ \ \ \ \ \ $d_{i,k} := d_{i,k} - qd_{j,k}$}


\RETURN the $n\times n$ matrix $D$.

\end{algorithmic}\label{algo:modified-Herimite-reduction}
\end{algorithm}

If Algorithm \ref{algo:modified-Herimite-reduction} output $D$ for
an $n\times (n-1)$ matrix $H$,  we say that $DH$ is the modified
Hermite reduction of $H$ and that $D$ is the reducing matrix of $H$.
This reduction develops the left multiplying modified Hermite
reducing matrix $D$.

Hermite reduction is also presented in \cite{FBA1999}, and is
equivalent to modified Hermite reduction  for a lower triangular
matrix $H$ with $h_{j,j} \neq 0$ (see \cite[Lemma 3]{FBA1999}). Both
the two equivalent reductions have the following properties:

1. The reducing matrix $D \in GL(n, \mathbb{Z})$.

2. For all $k > i$, the (modified) Hermite reduced matrix $H' =
(h'_{i,j}) = DH$ satisfies $|h'_{k,i}| \leq |h'_{i,i}|/ 2 =
|h_{i,i}|/ 2$.

In order that the reduced and reducing matrices of $H_{X}
\in\mathbb{R}^{n\times(n-t)}$ satisfy the two properties above, we
need the following generalized Hermite reduction. 

\begin{algorithm}[H]
\caption{(Generalized Hermite Reduction).}
\begin{algorithmic}[1]
\REQUIRE a lower trapezoidal matrix $H=(h_{i,j})\in
\mathbb{R}^{n\times (n-t)}$ with $h_{j,j} \neq 0$.

\ENSURE a reducing matrix $D$ of $H$.

\STATE {$D := I_n$}

\STATE {\textbf{for} $i$ from $2$ to $n$
\textbf{do}}\label{algostep:ghr-hermite-reduction-start}


\STATE {\ \ \ \ \textbf{if} $i\leq n-t+1$ \textbf{then} $temp :=
i-1$\\\ \ \ \ \textbf{else} $temp := n-t$}


\STATE {\ \ \ \ \textbf{for} $j$ from $temp$ by $-1$ to $1$
\textbf{do}}

\STATE {\ \ \ \ \ \ \ \ $q := \lfloor h_{i,j}/h_{j,j}\rceil$}

\STATE {\ \ \ \ \ \ \ \ \textbf{for} $k$ from $1$ to $n$
\textbf{do}}

\STATE {\ \ \ \ \ \ \ \ \ \ \ \ $d_{i,k} := d_{i,k} - qd_{j,k}$}








\STATE {\textbf{for} every two integers $s_1,  s_2 \in \{n-t+1,
\cdots, n\}$ satisfying $s_1 < s_2$, $h_{s_1, n-t} = 0$ and $h_{s_2,
n-t} \neq 0$
\textbf{do}}\label{algostep:generalized-2-hermite-reduction-start}

\STATE {\ \ \ \ exchange the $s_1$-th row and the $s_2$-th row of
$D$.} \label{algostep:generalized-2-hermite-reduction-end}

\RETURN the $n\times n$ matrix $D$.

\end{algorithmic}\label{algo:generalized-Herimite-reduction}
\end{algorithm}


If Algorithm \ref{algo:generalized-Herimite-reduction} output $D$
for an $n\times (n-t)$ matrix $H$, we call $DH$ the generalized
Hermite reduction of $H$ and $D$ the reducing matrix of $H$.
Obviously, generalized Hermite reduction is equivalent to modified
Hermite reduction when $t=1$. In addition, we can easily check that
generalized Hermite reduction remains the two properties mentioned
above.


\begin{rem}\label{rem:generalized-Hermite-reduction}
 There are two main differences between
(modified) Hermite reduction and  generalized Hermite reduction.
Firstly, the last $t-1$ rows of $H$ will also be reduced by the
first $n-t$ rows of $H$ in generalized Hermite reduction, while
(modified) Hermite reduction can not do so. Secondly, generalized
Hermite reduction exchanges the $s_1$-th row and the $s_2$-th row of
$D$ if $s_1 < s_2$, $h_{s_1, n-t} = 0$ and $h_{s_2, n-t} \neq 0$
(from Step \ref{algostep:generalized-2-hermite-reduction-start} to
Step \ref{algostep:generalized-2-hermite-reduction-end}). This
implies that if $h_{n-t+1, n-t}= 0$ after  generalized Hermite
reduction then $h_{n-t+2, n-t} = \cdots = h_{n, n-t} = 0$. This
property plays an important role in the proof of Lemma
\ref{lem:Hn,n-2=0}.
\end{rem}

\section{The SIRD Algorithm}\label{sec:The SIRD algorithm}
\subsection{The Description of SIRD}
\label{subsec:The algorithm description} Using the hyperplane matrix
constructing method and generalize Hermite reduction in the previous
section we can get a simultaneous integer relation detecting
algorithm SIRD.

\begin{algorithm}[t!]
 \caption{(The SIRD Algorithm).}
\begin{algorithmic}[1]
\REQUIRE $(\mathbf{x_1}, \cdots, \mathbf{x_t}) =
X\in\mathbb{R}^{n\times t}$ 
satisfying (\ref{eq:suppose-on-x})

\ENSURE either output an SIR for $X$ or give a lower bound on
$\lambda(X)$.

\STATE {\em Initiation.} Compute the hyperplane matrix $H_{X}$, set
$H := H_{X}$, $ B := I_n$.

\STATE {\em Reduction.} Call Algorithm
\ref{algo:generalized-Herimite-reduction} to reduce $H_{X}$
producing the reducing matrix $D\in GL(n, \mathbb{Z})$. Set $X :=
XD^{-1}, H := DH,  B := BD^{-1}$.

\LOOP

\STATE\label{algostep:exchange} {\em Exchange.} Let $H = (h_{i,j})$.
Choose an integer $r$ such that $\gamma^{r}|h_{r,r}| \geq
\gamma^{i}|h_{i,i}|$ for $1\leq i\leq n - t$, where $\gamma >
2/\sqrt{3}$. Define the permutation matrix $R$ to be the identity
matrix with the $r$ and $r+1$~rows exchanged. Update $X := XR, H :=
RH,  B := BR$.

\STATE\label{algostep:corner} {\em Corner.} Let
\begin{equation}\label{eq:alph-beta-lambda-delta}
\begin{array}{ll}
  \alpha := h_{r,r}, & \beta := h_{r+1,r}, \\
  \lambda := h_{r+1, r+1}, & \delta := \sqrt{\beta^2 +
\lambda^2}.
\end{array}
\end{equation} Let $Q := I_{n-t}$. If
$r < n - t$,
 then let the submatrix of $Q$ consisting of the $r$-th and
$(r + 1)$-th rows of columns $r$ and $r + 1$ be
$\left (\begin{matrix} \beta/\delta & -\lambda/\delta\\
\lambda/\delta&\beta/\delta
\end{matrix}
\right )$.
Update $H := HQ$.

\STATE\label{algostep:reduction} {\em Reduction.} Call Algorithm
\ref{algo:generalized-Herimite-reduction} to reduce $H_{X}$
producing $D$. Update $X := XD^{-1}, H := DH,  B := BD^{-1}$.

\STATE Compute $G := 1/\max_{1\leq j\leq n-t}\lvert h_{j,j}\rvert$.
Then there  exists no SIR whose Euclidean norm is less than $G$.
\label{algostep:bound-of-possible-relations}

\STATE{\textbf{if} $\mathbf{x_j} = 0$ for some $1 \leq j \leq n$, or
$h_{n-t,n-t}=0$ \textbf{then}}

\STATE {\ \ \ \ \textbf{return} the corresponding SIR for $X$.}


\ENDLOOP
\end{algorithmic}\label{algo:Simultaneous-relation}
\end{algorithm}

\subsection{Analysis of SIRD}
Let $H(k)$ be the result after  $k$ iterations of SIRD.

Why do we set the parameter $\gamma > 2/\sqrt{3}$ at Step
\ref{algostep:exchange}?  Suppose the $r$ chosen in Step
\ref{algostep:exchange} is not $n - t$. In this case we let $\alpha,
\beta, \lambda, \delta$ be as in (\ref{eq:alph-beta-lambda-delta}).
Then
\[
\left (\begin{matrix} \alpha&0\\
\beta&\lambda
\end{matrix}
\right )
\]
is the submatrix of $H(k-1)$ consisting of the $r$ and $r + 1$ rows
of columns $r$ and $r + 1$, where $r < n - t$.   After Step
\ref{algostep:exchange} has been performed $\lambda$ may not be
zero, which makes that $H$ is not lower trapezoidal.  
After Step \ref{algostep:corner} the result is
\begin{equation}\label{eq:exchange}
\left (\begin{matrix}\beta&\lambda  \\
\alpha&0
\end{matrix}
\right )
\left (\begin{matrix} \beta/\delta & -\lambda/\delta\\
\lambda/\delta&\beta/\delta
\end{matrix}
\right )  =
\left (\begin{matrix} \delta & 0\\
\alpha \beta/\delta&-\alpha\lambda/\delta
\end{matrix}
\right ).
\end{equation}
Since $r$ is chosen such that $\gamma^r|h_{r,r}(k - 1)|$ is as large
as possible, and $r < n - t$ we have $|h_{r+1,r+1}(k - 1)| \leq
\frac{1}{\gamma}|h_{r,r}(k - 1)|$, hence $|\lambda| \leq
\frac{1}{\gamma}|\alpha|$. From the property of generalized Hermite
reduction 
we have that $|\beta| \leq
\frac{1}{2}|\alpha|$, which then gives
\begin{equation}\label{eq:delta}
\left|\frac{h_{r,r(k)}}{h_{r,r(k-1)}}\right| =
\left|\frac{\delta}{\alpha}\right| = \sqrt{\frac{\beta^2 +
\lambda^2}{\alpha^2}} \leq  \sqrt{\frac{1}{4} + \frac{1}{\gamma^2}}.
\end{equation}
Thus $|h_{r,r}|$ is reduced as long as $\sqrt{\frac{1}{4} +
\frac{1}{\gamma^2}} < 1$, i.e. $\gamma > 2/\sqrt{3}$. As was pointed
out by Borwein (see \cite
{Bor2002}), although this
increases $h_{r+1,r+1}$, this is not a significant problem. At each
step we force the larger diagonal elements of $H$ toward
$h_{n-t,n-t}$, where their size can be reduced by at least a factor
of $2$ when $r = n-t$.

As a matter of fact, the parameter $\gamma$ can be freely chosen in
the open interval $(2/\sqrt{3}, +\infty)$.


\begin{lem}\label{lem:Hn,n-2=0}
If $h_{j,j}(k) = 0$ for some $1 \leq j \leq n - t$ and no smaller
$k$, then $j = n - t$ and an SIR for $X$ must appear as a column of
the matrix $B$.
\end{lem}
\begin{proof}
By the hypothesis on $k$ we know that all diagonal elements of $H(k
- 1)$ are not zero. Now, suppose the $r$ chosen in Step
\ref{algostep:exchange} is not $n-t$. Since generalized Hermite
reduction does not introduce any new zeros on the diagonal, and from
the analysis of Step \ref{algostep:exchange} and Step
\ref{algostep:corner} above, we have that no diagonal element of
$H(k)$ is zero. This contradicts the hypothesis on $k$ and our
assumption that $r < n - t$ was false. Thus we have $r = n - t$
after the $(k - 1)$-th iteration has been completed.

Next we show that there must be an SIR for $X$ appeared as a column
of the matrix $B$. We have $X^TH_{X} = \mathbf{0}$ from Lemma
\ref{lem:property-of-Hx} and hence that  $\mathbf{0} =
X^TBB^{-1}H_{X} = X^TBB^{-1}H_{X}Q = X^TBH(k - 1)$, where $Q$ is an
appropriate orthogonal $(n - t)\times(n - t)$ matrix. Let
$(\mathbf{z_1}, \cdots, \mathbf{z_t})^T = X^TB$, where
$\mathbf{z_i}=(z_{i,1},\cdots,z_{i,n})^T$. Then
 \[
 \begin{split}
&        \left
(\begin{matrix}0&\cdots&0\\
\vdots&\ddots &\vdots\\
0&\cdots&0\end{matrix}
 \right ) = X^TBH(k - 1) = \left (\begin{matrix}\mathbf{z_1}^T\\\vdots\\\mathbf{z_t}^T\end{matrix}
 \right )H(k - 1)\\&= \left
 (\begin{matrix}\cdots,&\sum_{k=n-t}^{n}z_{1,k}h_{k,n-t}(k-1)\\\cdots,&\cdots
\\\cdots,&\sum_{k=n-t}^{n}z_{t,k}h_{k,n-t}(k-1)\end{matrix}
 \right )\\
 &= \left (\begin{matrix}\cdots,&z_{1,n-t}h_{n-t, n-t}(k - 1)\\\cdots,&\cdots\\
 \cdots,&z_{t,n-t}h_{n-t, n-t}(k - 1)
 \end{matrix}
 \right ).
 \end{split}
 \]
We know $h_{n-t+1, n-t}(k - 1) =0$ and $h_{n-t, n-t}(k - 1) \neq 0$
from $h_{n-t, n-t}(k) = 0$. From Remark
\ref{rem:generalized-Hermite-reduction} and $h_{n-t+1, n-t}(k - 1)
=0$ we have $h_{n-t+2, n-t}(k - 1) = \cdots = h_{n, n-t}(k - 1) = 0$
which implies the last equality. Since $h_{n-t, n-t}(k - 1) \neq 0$,
it follows that $z_{1, n-t} = \cdots = z_{t, n-t} = 0$. Thus the $(n
- t)$-th column of $B$ is an SIR for $X$.
\end{proof}

From Theorem \ref{thm:lower-bound-of-simultaneous-relations} and
Lemma \ref{lem:Hn,n-2=0}, the correctness of SIRD has been proved.
Moreover, we have

\begin{thm}\label{thm:upper-bound-for-SIRs}
Let $\lambda(X)$ be the least Euclidean norm of any SIR for $X$. Let
$\mathbf{m}$  be an SIR detected by SIRD. Then
$\|\mathbf{m}\|_2\leq\gamma^{n-t-1}\lambda(X)$ for all
$\gamma>2/\sqrt{3}$.
\end{thm}
\begin{proof}
Assume $r=n-t$ with $h_{n-t,n-t}(k)\neq 0$ and $h_{n-t,n-t}(k+1)=0$
at the $k$-th iteration of SIRD. Then from Theorem
\ref{thm:lower-bound-of-simultaneous-relations} and the exchange
rule of SIRD we have
\[
\lambda(X)\geq 1/\max_{1\leq i\leq
n-t}|h_{i,i}(k)|\geq\gamma^{t+1-n}/|h_{n-t,n-t}(k)|.
\] At this time, $\|\mathbf{m}\|_2=1/|h_{n-t,n-t}(k)|$ holds from
the same strategy in the proof of Lemma 10 in \cite{FBA1999}.
\end{proof}
\begin{defn}[the $\Pi$ function]\label{def:Pi-function}
For the $k$-th iteration in SIRD, define
\[
\Pi(k) = \prod_{1\leq j\leq n-t} \min\left\{\gamma^{n-t}\lambda(X),
\frac{1}{\left|h_{j,j}(k)\right|}\right\}^{n-j}.
\]
\end{defn}

The routine of analyzing the number of iterations in \cite{FBA1999}
can be carried over here with redefining the $\Pi$ function as
above. So we state the following lemma directly without proof.

\begin{lem}\label{lem:Pi-function}
For $k>1$ we have

1. $\left(\gamma^{n-t}\lambda(X)\right)^{\left[\left(
                                                          \begin{matrix}
                                                            n \\
                                                            2 \\
                                                          \end{matrix}
                                                        \right)-\left(
                                                          \begin{matrix}
                                                            t \\
                                                            2 \\
                                                          \end{matrix}
                                                        \right)
\right]}\geq\Pi(k)\geq 1$, where $\lambda(X)$ is the least norm of
SIRs for $X$.

2. $\Pi(k)\geq \sqrt{\frac{4\gamma^2}{\gamma^2+4}}\, \Pi(k-1)$.
\end{lem}



From this lemma, it follows that the $\Pi$ function is increasing
with respect to $k$ and has an upper bound for a fixed
$\gamma\in(2/\sqrt{3},+\infty)$. Thus we have

\begin{thm}\label{thm:the-number-of-iterations}
If $X\in\mathbb{R}^{n\times t}$ has SIRs, then the number of
iterations such that SIRD finds an SIR for $X$ will be no more than
\[
\left[\left(\begin{matrix}
              n \\
              2
            \end{matrix}
\right)-\left(\begin{matrix}
              t \\
              2
            \end{matrix}
\right)\right] \frac{\log(\gamma^{n-t}\lambda(X))}{\frac{1}{2}
\log\left(\frac{4\gamma^2}{\gamma^2+4}\right)}.
\]
\end{thm}

\begin{proof}
From Definition \ref{def:Pi-function} we can infer $\Pi(0)\geq 1$.
And by Lemma \ref{lem:Pi-function} we know that \[\Pi(k)\geq
\left(\sqrt{\frac{4\gamma^2}{\gamma^2+4}}\, \right)^k.\] Solving $k$
from this inequality gives the conclusion, as was to be shown.
\end{proof}

\begin{cor}\label{thm:time-complexity-of-simultaneous-relation}
If $X\in\mathbb{R}^{n\times t}$ has SIRs, then there exists a
$\gamma$ such that SIRD will find an SIR for $X$ in polynomial time
$\mathcal{O}(n^4+n^3\log\lambda(X))$.
\end{cor}

\begin{proof}
Let $\gamma = 2$. Then SIRD will construct an SIR for $X$  in no
more than
\[
(n-t)^2(n+t-1)+(n-t)(n+t-1)\log\lambda(X)
\]
iterations. SIRD takes $\mathcal{O}(n-t)$ exact arithmetic
operations per iteration, and hence that
$\mathcal{O}((n-t)^4+(n-t)^3\log\lambda(X))$ exact arithmetic
operations is enough to produce an SIR for $X$. Since $t<n$, the
proof is complete.
\end{proof}

\begin{rem}\label{rem:simultaneous-relation-algorithm}
From this corollary, we can claim that our detecting algorithm
always return an SIR for $X$ if one exists. Additionally, SIRD will
produce lower bound on the Euclidean norm of any possible SIRs for
$X$ (Theorem \ref{thm:lower-bound-of-simultaneous-relations}). Thus
SIRD can be used to prove that there are no SIRs for $X$ of norm
less than a given size.
\end{rem}

\begin{rem}
PSLQ may be viewed as a particular case of SIRD when $t = 1$.
Similarly with PSLQ, SIRD can be easily generalized to complex field
with $\gamma>\sqrt{2}$ such that the  outputs are in Gaussian
integer ring and all conclusions mentioned above hold with
corresponding modifications.
\end{rem}
\begin{rem}
Moreover, SIRD can also be applied to detect an integer relation in
$\mathbb{Z}^n$ for a given complex vector. For example, suppose
$\mathbf{z} = \mathbf{x} + \mathbf{y}I$ in $\mathbb{C}^n$ with
vector components $\mathbf{x}, \mathbf{y} \in \mathbb{R}^n$ where $I
= \sqrt{-1}$. Then SIRD can give an SIR $\mathbf{m}$ for
$(\mathbf{x},\mathbf{y})$, and hence that
$\mathbf{m}\in\mathbb{Z}^n$ is an integer relation for $\mathbf{z}$,
but PSLQ only can give a Gaussian integer relation in
$\mathbb{Z}[I]^n$. 
 This is one of
the biggest differences between SIRD and PSLQ. Furthermore, 
the matrix reducing method in SIRD is generalized Hermite reduction,
which avoids LLL-type reduction. This is  a difference not only
between SIRD and HJLS, but also between SIRD and PSLQ because that
(modified) Hermite reduction is not suitable to detect 
SIRs any more. And just the generalized Hermite reduction guarantees
the correctness of SIRD.
\end{rem}


\section{Performance Results}\label{sec:Performance results}
In theory, the costs of SIRD and the HJLS simultaneous relations
algorithm (see \cite[section 5]{HJL1989}) are the same as in
Corollary \ref{thm:time-complexity-of-simultaneous-relation} in the
worst case, whereas in practice  SIRD usually needs fewer
iterations. For $\mathbf{v_1} = (11,27,31)^T$ and $\mathbf{v_2} =
(1,2,3)^T$, HJLS outputs $(19, -2, -5)^T$ after 5 iterations while
SIRD outputs $(-19, 2, 5)^T$ after only 2 iterations.

\begin{table}[H]\centering
\begin{tabular}{||c|c|c|c|c|c|c|c|}
  \hline
No.&  $n$  & $itr_{HJLS}$ & $itr_{SIRD}$ & $t_{HJLS}$ & $t_{SIRD}$
 \\\hline\hline
  1&4&15&12&\ \ 0.047&0.\ \ \  \\
 2&4&13&9&\ \ 0.171&\ \ 0.016\\
 3&4&21&19&\ \ 0.062&\ \ 0.015\\
 4&5&25&20&\ \ 0.110&\ \ 0.016\\
 5&5&27&43&\ \ 0.125&\ \ 0.016\\
 6& 5 &   21 & 14 & \ \ 0.110 & \ \ 0.032 \\\hline
7& 30 &  51 & 21 & \ \ 1.703 & \ \ 0.422 \\
  8&54 & 34 & 9 & \ \ 5.625 & \ \ 1.265 \\
   9&79  & 34 & 40 & \ 14.157 & \ \ 4.422 \\
  10&97 &  37 & 5 & \ 23.860 & \ \ 5.375 \\
  11&128 &  45 & 6 & \ 49.657 & \ 11.141 \\
  12&149 &   29 & 14 & \ 76.797 & \ 18.063 \\
   13&173 & 26 & 2 & 114.140 & \ 25.000 \\
  14&192 &  29 & 2 & 153.078 & \ 33.641 \\
  15&278& 28 & 8 & 440.781 & 102.860  \\
   16&290 & 35 & 6 & 500.562 & 118.578  \\
   17&293 & 23 & 7 & 512.796 & 123.265  \\
   18&305 & 22 & 4 & 581.844 & 137.672  \\
   19&316 & 19 & 3 & 649.032 & 147.796  \\
  20&325 & 18 & 2 & 716.094 & 159.813  \\
  \hline
\end{tabular}\caption{Comparison of performance results for HJLS and SIRD }\label{tab:performance-of-SIRD}
\end{table}

 Both the SIRD algorithm and the HJLS simultaneous relations
algorithm when $t=2$, i.e. detecting an SIR for two vectors, were
implemented in \emph{Maple} 13 by the first author.  The tests were
run on AMD Athlon$^{\tiny\mbox{TM}}$ 7750 processor (2.70 GHz) with
2GB main memory.

The purpose of the trials in Table \ref{tab:performance-of-SIRD} is
to compare the performances of HJLS and SIRD.  $n$ in Table
\ref{tab:performance-of-SIRD} gives the dimension of the relation
vector. $itr_{HJLS}$ and $itr_{SIRD}$ are the numbers of iterations
of HJLS and SIRD respectively. The columns headed $t_{HJLS}$ and
$t_{SIRD}$ give the CPU run time respectively of the two algorithms
in seconds. 

The 20 trials in Table \ref{tab:performance-of-SIRD} were
constructed by \emph{Maple}'s pseudo random number generator. The
first $6$ trials are for low dimension, and  others for higher
dimension. The results show that SIRD appears to be more effective
than HJLS. In $18$ out of $20$ trials, the number of iterations of
SIRD is less than that of HJLS. It is still true that SIRD usually
needs fewer iterations than  HJLS for more tests. This leads that
the running time of SIRD is much less than HJLS.  With $n$
increasing, the difference between the efficiency of SIRD and HJLS
is increasingly notable. On average, the SIRD running time is about
$26.7\%$ of the running time of HJLS. All these results are obtained
under the condition that $\gamma=2/\sqrt{3}+10^{-14}$.

The \emph{Maple} implementation and more tests are available from
{\tiny
\url{http://cid-5dbb16a211c63a9b.skydrive.live.com/self.aspx/.Public/sird.rar}.
}

\section{An Application}
\label{sec:finding-minimal-polynomial} Any SIR detecting algorithm
intervenes in many fields of application, such as Diophantine
approximating, numerical constants relations finding, etc. In this
section, we discuss how to find the minimal polynomial of a complex
algebraic number from its an approximation by using SIRD.
\subsection{The MPF Algorithm}
We say that a complex number $\alpha$ is an algebraic number if
$\alpha$ is a root of a non-zero polynomial in one variable with
integer coefficients. The minimal polynomial of $\alpha$ is the
unique primitive polynomial $p(x) \in \mathbb{Z}[x]$ of least degree
such that $p(\alpha)  = 0$. The degree and  height of $\alpha$ are
the degree and  height of its minimal polynomial $p(x)$
respectively.

In this section, let $\alpha = a + bI \in \mathbb{C}$ be an
algebraic number with degree at most $n$, height at most $H$
, where $I = \sqrt{-1}$. Suppose we are given an approximation
$\bar{\alpha}$ to  $\alpha$ such that
\begin{equation}\label{eq:epsilon}\max_{1\leq i\leq
n}|\alpha^{i}-\bar{\alpha}^{i}| < \epsilon.\end{equation} Is it
possible to infer the minimal polynomial from the approximation?
Computer algebra system \emph{Maple} has an LLL-based procedure,
\texttt{PolynomialTools:-MinimalPolynomial()}, for finding the
minimal polynomial of an algebraic number from its an approximation,
whose
basic idea is from \cite{Sch1984, KLL1984, KLL1988}. 
Applying SIRD, we shall give another affirmative answer, the
following MPF algorithm, to the question above.

\begin{algorithm}[H]
\caption{(The MPF Algorithm).}
\begin{algorithmic}[1]
\REQUIRE an approximation $\bar{\alpha}$ to $\alpha$ satisfying
(\ref{eq:epsilon}), a degree bound $n$, and a height bound $H$,
$\epsilon$ satisfying (\ref{eq:error-controlling})

\ENSURE 
the minimal polynomial of $\alpha$.

\WHILE{  $2\leq i\leq n$}\label{algostep: circulating}

\STATE { $\mathbf{v} := (1, \bar{\alpha}, \cdots,
\bar{\alpha}^i)^T$}


\STATE { Call SIRD with $\gamma=2$ producing an integer relation
$\mathbf{p_i}=(p_0, p_1, \cdots, p_i)^T$ for
$\mathbf{v}$\\$p_i:=$the primitive part of $\sum_{j=0}^{i}
 p_jx^{j}$
}\label{algostep:call-SIRD}







\STATE {\textbf{if} height$(p_i)>2^{n-2}\sqrt{n+1}H$ 
\textbf{then}}\label{algostep:if height(p)>H}

\STATE { \ \ \ \ $i:=i+1$; \textbf{goto} Step \ref{algostep:
circulating}}

\STATE { \textbf{else}  \textbf{return} $p_i$}







\ENDWHILE
\end{algorithmic}\label{algo:minimal-polynomial}
\end{algorithm}

\begin{rem}
At Step \ref{algostep:call-SIRD} of MPF, $\mathbf{p_i}$ is an SIR
for $\mathbf{v_1} = (1, \mbox{Re}(\bar{\alpha}), \cdots,
\mbox{Re}(\bar{\alpha}^i))^T$ and $\mathbf{v_2} = (0,
\mbox{Im}(\bar{\alpha}), \cdots, \mbox{Im}(\bar{\alpha}^i))^T$ when
 $\mbox{Im}(\alpha)\neq 0$.
\end{rem}

\subsection{Error Controlling}
\label{subsec:Finding the Minimal Polynomial Using Float-point
Arithmetic}



 The main idea of our minimal
polynomial finding (MPF) algorithm to determine the minimal
polynomial of an algebraic number from its an approximation is as
follows: We try the value of  $i=2,\cdots,n$ in order. With $i$
fixed, we call SIRD for detecting an exact integer relation
$\mathbf{p_i}=(p_0,p_1,\cdots,p_i)^T$ for $\mathbf{v}=(1,
\bar{\alpha}, \cdots, \bar{\alpha}^i)^T$. 
Then
$p_i(x)=\sum_{j=0}^ip_jx^j$ satisfies $p_i(\bar{\alpha})=0$,
however, from which we can not decide whether $p_i(\alpha)$ is $0$
or not. Hence the most important problem is how to choose an
appropriate $\epsilon$ in (\ref{eq:epsilon}) such that
$p_i(\bar{\alpha})=0$ implies $p_i(\alpha)=0$.
Before describing it
in detail, we consider the following example.

\begin{exmp}\label{exmp:x^2-4x+7}
Let $\alpha = 2 + \sqrt{3}I$. We know that the minimal polynomial of
$\alpha$ in $\mathbb{Z}[x]$ is $7-4x+x^2$. Let $\bar{\alpha} = 2.000
+ 1.732 I$ be the approximation to $\alpha$ with four significant
digits. Hence $\mathbf{v_1} = (1., 2., 1.)^T$, $\mathbf{v_2} = (0.,
1.732, 6.928)^T$. Feeding SIRD $\mathbf{v_1}$, $\mathbf{v_2}$ as its
input vectors gives an SIR for $\mathbf{v_1}$, $\mathbf{v_2}$ after
$2$ iterations. The corresponding matrices $B$ are
\[
 \left( \begin {array}{ccc} 2&1&0\\\noalign{\medskip}-1&-1&0
\\\noalign{\medskip}0&0&1\end {array} \right),
\left( \begin {array}{ccc}
7&0&2\\\noalign{\medskip}-4&0&-1\\\noalign{\medskip}1&1&0\end
{array} \right).
\]
It is obvious that the first column of the latter one
is an SIR for $\mathbf{v_1}$ and $\mathbf{v_2}$, and corresponds to
the coefficients of the minimal polynomial of $\alpha$. However, if
we take only $3$ significant digits for the same data, after $3$
iterations SIRD outputs $(1213,-693,173)^T$, which is an SIR for
$(1., 2., 1.)^T$ and $(0., 1.73, 6.93)^T$, but does not correspond
to the coefficients of the minimal polynomial of $\alpha$. For this
reason, we have to appropriately control the error such that the
output of MPF is correct.
\end{exmp}

%



\begin{lem}\label{lem:approximation-of-f}
Let $f$ be a polynomial in $\mathbb{Z}[x]$ of degree $n$. If
$\max_{1\leq i\leq n}|\alpha^{i}-\bar{\alpha}^{i}| < \epsilon$, then
$     |f(\alpha)-f(\bar{\alpha})| \leq  \epsilon \cdot n \cdot
     \mbox{height}(f)$.
\end{lem}

\begin{defn}[Mahler measure]
For any polynomial $g=\sum_{i=0}^{m}g_{i}x^{i}\in\mathbb{Z}[x]$ of
degree $m$ with the complex roots $z_{1},z_{2},\dots,z_{m}$ we
define the Mahler measure $M(g)$ by
\[
M(g)=|g_{m}|\prod_{j=1}^{m}\max\{1, |z_{j}| \}.
\]
The Mahler measure of an algebraic number $\alpha$ is defined to be
the measure of its minimal polynomial.
\end{defn}

\begin{lem}(see \cite[Lemma 3]{MW1978})\label{lem:Mignotte}
Let $\alpha_{1}, \cdots, \alpha_{q}$ be algebraic numbers of exact
degree of $d_{1}, \cdots, d_{q}$ respectively. Define
$D=[\mathbb{Q}(\alpha_{1}, \cdots, \alpha_{q}):\mathbb{Q}]$. Let $P
\in \mathbb{Z}[x_{1}, \cdots, x_{q}]$ have degree at most $N_{h}$ in
$x_{h}$ ($1\leq h\leq q$). If $P(\alpha_{1}, \cdots, \alpha_{q})
\neq  0$, then
\[
|P(\alpha_{1},\dots,\alpha_{q})| \geq  \|
P\|_{1}^{1-D}\prod_{h=1}^{q}M(\alpha_{h})^{-DN_{h}/d_{h}},
\]
where $M(\alpha)$ is the Mahler measure of $\alpha$.
\end{lem}


This lemma gives a lower bound on $|P(\alpha_{1}, \dots,
\alpha_{q})|$ if $P(\alpha_{1},\\\cdots, \alpha_{q}) \neq  0$ for an
arbitrary multivariate polynomial $P\in \mathbb{Z}[x_{1}, \cdots,
x_{q}]$. If we apply it to $g(x) = \sum_{i=0}^m g_i x^i$ in
$\mathbb{Z}[x]$, then we have
\begin{cor}\label{cor:g(alpha)not equal to 0}
Let $\alpha$ be an algebraic number with exact degree $n_0$ and
$g(x) = \sum_{i=0}^m g_i x^i \in \mathbb{Z}[x]$. Suppose both
$\mbox{height}(g)$ and $\mbox{height}(\alpha)$ are $\leq H$. If
$g(\alpha) \neq 0$, then
\[|g(\alpha)|  \geq (m +
1)^{-(n_0-1)}\cdot(n_0+1)^{-\frac{m}{2}}\cdot H^{-(m+n_0-1)},\]
where $\mbox{height}(\alpha)$ is the height of $\alpha$'s minimal
polynomial.
\end{cor}

\begin{proof}
For $f(x) \in \mathbb{Z}[x]$ with degree $n$, we
have 
Landau's inequality: $M(f) \leq \|f\|_2$ (e.g. see \cite[p.
154]{GG1999}), $\mbox{height}(f) \leq \|f\|_1 \leq
(n+1)\mbox{height}(f)$, and $\mbox{height}(f) \leq \|f\|_2 \leq
\sqrt{n+1}\mbox{height}(f)$.
This corollary  easily follows from Lemma \ref{lem:Mignotte} and the
three facts above.
\end{proof}

Next we investigate how to choose $\epsilon$ to enable MPF to
correctly return the minimal polynomial of $\alpha$ from
$\bar{\alpha}$. We denote the exact degree of $\alpha$ by $n_0(\leq
n)$. For $2\leq i\leq n$, Step \ref{algostep:call-SIRD} in MPF gives
a polynomial $p_i\in\mathbb{Z}[x]$ with degree $\leq i$ such that
$p_i(\bar{\alpha})=0$.
From Corollary \ref{cor:g(alpha)not equal to
0} we know that if $p_i(\alpha)\neq 0$, then
\begin{equation}\label{eq:p_i(alpha)<>0} |p_i(\alpha)| \geq (i +
1)^{(1-n_0)}\cdot(n_0+1)^{-\frac{i}{2}}\cdot H^{(1-i-n_0)}\geq M ,
\end{equation}
where $M = (n + 1)^{-(\frac{3}{2}n-1)}\cdot
H^{-(2n-1)}$. 

\begin{thm}
Let $\alpha$, $\bar{\alpha}$ and $M$ be as above, 
and $p$ a polynomial in $\mathbb{Z}[x]$ with degree $\leq n$ and
height $\leq H$. Then there exist some $\epsilon$ such that
$|p(\bar{\alpha})| =0$ implies $p(\alpha) = 0$.
\end{thm}

\begin{proof}

Set $\deg(p)=i(\leq n)$. From Lemma \ref{lem:approximation-of-f} we
have $|p(\alpha)|=|p(\alpha)-p(\bar{\alpha})|\leq \epsilon\cdot
i\cdot H\leq \epsilon\cdot n\cdot H$. Thus if 
 $\epsilon<\frac{M}{nH}$,
then  $|p(\alpha)|<M$. From (\ref{eq:p_i(alpha)<>0}) it follows that
$p(\alpha)=0$.
\end{proof}

If we substitute $2^{n-2}\sqrt{n+1}H$ for $H$, we have

\begin{cor}\label{cor:epsilon-control}
Let $\alpha$ and $\bar{\alpha}$  be as above and
\begin{equation}\label{eq:error-controlling}
\epsilon<2^{-2n^2+4n}(n+1)^{-\frac{5}{2}n}H^{-2n}.
\end{equation} Then for $i$ from $1$ to $n$, an integer relation for
$(1,\bar{\alpha},\cdots,\bar{\alpha}^i)^T$ with height $\leq
2^{n-2}\sqrt{n+1}H$ is also for $(1,\alpha,\cdots,\alpha^i)^T$.
\end{cor}

\subsection{Correctness and Cost of MPF}

Assume that the degree of $\alpha$ is $n_0$ and that $\epsilon$
satisfies (\ref{eq:error-controlling}). When $2\leq i<n_0$, there
exists no relation for $(1,\alpha,\cdots,\alpha^i)$, which, combined
with Corollary \ref{cor:epsilon-control}, means that $p_i(x)$ must
satisfy the condition in Step \ref{algostep:if height(p)>H} of MPF
and then go into next iteration. When $i=n_0\,(<n)$, we know that
the coefficients of the minimal polynomial of $\alpha$ form an
integer relation for $(1,\alpha,\cdots,\alpha^{n_0})$, whose height
$\leq H$, hence Euclidean norm $\leq\sqrt{n_0+1}H$. This implies
that $(1,\bar{\alpha}, \cdots,\bar{\alpha}^{n_0})$ has also an
integer relation with Euclidean norm $\leq\sqrt{n_0+1}H$. From
Theorem \ref{thm:upper-bound-for-SIRs} we know that the height of
the relation SIRD detected will $\leq 2^{n-2}\sqrt{n+1}H$. Thus the
relation detected by SIRD when $i=n_0$ will never satisfy the
condition in Step \ref{algostep:if height(p)>H}  and corresponds an
integral multiple of the minimal polynomial of $\alpha$. Hence the
correctness of MPF follows.

From  (\ref{eq:error-controlling}) we have $\log \epsilon
 \in \mathcal{O}(n^2 + n\log H)$. Thus we can give
another answer to  Blum's and Zhang's question without using LLL
lattice reduction algorithm.

\begin{thm}\label{thm:complexity-of-finding-algorithm-fpa}
Let $\alpha$ be an algebraic number and let $n$ and $H$ be upper
bounds of the degree and height of $\alpha$ respectively. Suppose we
are given an approximation $\bar{\alpha}$ to $\alpha$ such that
$\max_{1\leq i\leq n}|\alpha^{i}-\bar{\alpha}^{i}| < \epsilon$. Then
the minimal polynomial of $\alpha$ can be determined in $\mathcal
{O}(n^5 + n^4\log H))$ arithmetic operations on floating-point
numbers having $\mathcal{O}(n^2 + n\log H))$ bit-complexity.
\end{thm}

\begin{table}[H]
\begin{tabular}{||c|c|c|}
  \hline
   & Digits & Complexity \\\hline
  KLL\cite{KLL1988} & $\mathcal{O}(n^2+n\log H)$ & $\mathcal
{O}(n^5 + n^4\log H)$ \\\hline
  Just\cite{Jus1989} & $\mathcal
{O}(n^2 + n^2\log H)$ & $\mathcal {O}(n^8\log n + n^8\log H)$
\\\hline
  QFCZ\cite{QFC2009} & $\mathcal{O}(n^2 + n\log H)$ & ------ \\\hline
  MPF & $\mathcal{O}(n^2 + n\log H))$ & $\mathcal
{O}(n^5 + n^4\log H))$ \\
  \hline
\end{tabular}\caption{Comparison of different minimal polynomial finding
algorithms}\label{tab:comparison of MPFs}
\end{table}

Table \ref{tab:comparison of MPFs} gives a comparison of the digits
and complexity of 4 different minimal polynomial finding algorithms
in the worst case. 
Since the algorithm in \cite{QFC2009} can only find the minimal
polynomial of a real algebraic number, we don't compare the
complexity with it. It seems that a lower complexity can be achieved
by using some new type LLL algorithms, such as L$^2$ \cite{NS2005}
and H-LLL \cite{MSV2009}, but when we apply these new algorithms to
find the minimal polynomial we have to choose $\epsilon$ as in a
similar formula with (\ref{eq:error-controlling}). Thus multiple
precision arithmetic is inevitable.





\vspace{2 mm}

\emph{Example \ref{exmp:x^2-4x+7} (con.).} For $\alpha = 2 +
\sqrt{3} I$, its minimal polynomial $7-4x+x^2$. Set $n = 2$ and $H =
7$. Computing the error tolerance as in equation
(\ref{eq:error-controlling}) gives $\epsilon <583443^{-1}$.
Corollary \ref{cor:epsilon-control} implies that
$\lfloor-\log_{10}583443^{-1}\rfloor = 5$ correct decimal digits are
sufficient to guarantee the output is correct. This example also
illustrates that $\epsilon$ in (\ref{eq:error-controlling}) is only
a sufficient condition on error controlling, but not a necessary
one.



\section{Conclusion}\label{sec:conclude}

The number of iterations and the cost of SIRD algorithm are related
to the parameter $\gamma$. For $\mathbf{v}_1=(86,  6, 8, 673)^T$ and
$\mathbf{v}_2=(83, 5, 87, 91)^T$, if we choose $\gamma=1.16$ then
SIRD outputs $(-215, 402, 159, 22)^T$ after 12 iterations, however,
if we choose $\gamma=5$, SIRD outputs  $(93, 364, 93, -14)^T$ after
only 6 iterations. In future work we expect to find the best choice
for $\gamma$. Additionally, how to choose the digits such that SIRD
under floating-point arithmetic finds an exact SIR is also in our
interests.
%
Finally, we see that the MPF algorithm can be used to factor $f$ in
$\mathbb{Z}[x]$ like this: Solve an approximation root with accuracy
satisfying  equation (\ref{eq:error-controlling}), and  call MPF for
finding its minimal polynomial which corresponds an irreducible
factor of $f$, and then repeat the two steps until $f$ has been
factored completely.
It is symbolic-numeric  and different from traditional algorithms
based on Hensel lifting.\vspace{1 mm}

\textbf{Acknowledgements.} This research was partially supported by
the Knowledge Innovation Program of CAS 
(KJCX\\2-YW-S02) and the NSFC (10771205).

\balance
\end{document}